\documentclass{article} 
\usepackage{graphicx} 
\usepackage{mn}
\def\be{\begin{equation}} 
\def\ee{\end{equation}} 
\def\ba{\begin{eqnarray}} 
\def\ea{\end{eqnarray}} 
 
\def\th{$^{th}$} 
\begin{document} 

\onecolumn

\title[Bayesian joint estimation of non-Gaussianity and the power
spectrum]{Bayesian joint estimation of non-Gaussianity and the power
spectrum}
\author[Gra\c{c}a Rocha, Jo\~ao Magueijo, Mike Hobson, 
Anthony Lasenby]{Gra\c{c}a Rocha$^1$$^,$$^2$, Jo\~ao Magueijo$^3$, Mike Hobson$^4$, 
Anthony Lasenby$^4$\\
$^1$Centro de Astrof\'{\i}sica, Universidade do Porto,
Rua das Estrelas s/n, 4150-762 Porto, Portugal\\
$^2$Dept. of Physics, University of Oxford, 
Nuclear \& Astrophysics Laboratory,
Keble Road, Oxford OX1 3RH, U.K.\\ 
$^3$Theoretical Physics, The Blackett Laboratory, 
Imperial College, Prince Consort Rd., London, SW7 2BZ, U.K.\\
$^4$Astrophysics Group, Cavendish Laboratory, 
Madingley Road, Cambridge, CB3 0HE, U.K.\\
} 
\date{\today} 
\maketitle

\begin{abstract} 
We propose a rigorous, non-perturbative, Bayesian framework
which enables one jointly to test Gaussianity and estimate the power spectrum 
of CMB anisotropies. It makes use of the Hilbert space of an harmonic 
oscillator to set up an exact likelihood function, dependent on the power spectrum 
and on a set of parameters $\alpha_i$, which are zero for 
Gaussian processes. 
The latter can be expressed as series of cumulants; indeed they 
perturbatively reduce to cumulants. However they have the advantage 
that their variation is essentially unconstrained. Any truncation
(i.e.: finite set of $\alpha_i$) therefore still produces a proper 
distribution - something which cannot be said of the only other such 
tool on offer, the Edgeworth expansion. We apply our method to
Very Small Array (VSA) 
simulations based on signal Gaussianity, showing that our algorithm is 
indeed not biased. 
\end{abstract}

\section{Introduction} 
With recent dramatic improvements in observations \cite{boom,max} the  
cosmic microwave background (CMB) temperature anisotropies  
have become a gold mine of information in cosmology. In  
particular, the power spectrum $C_\ell$ of CMB fluctuations has become a 
popular meeting ground for theorists and observers.  On one hand measuring
the $C_\ell$  provides a concrete target for improving experimental
strategies. On another, well-founded theories predict that a system 
of peaks should be observed in the $C_\ell$, 
with a wealth of information encoded in their positions 
and heights. The recent unambiguous detection of the first of these peaks  
\cite{boom,max,toco} has caused a great deal of excitement  
(e.g. \cite{melch,dod,cont}).

Many reputable theories also predict that the  
temperature anisotropies form a  Gaussian random field
(even though there are notable exceptions 
\cite{salopek,mukh,peebles,joao}). 
That being the case, the power spectrum does indeed  
contain all the relevant information characterizing 
the random process under  
study. But how can we test the hypothesis of Gaussianity?  
Unfortunately the formalism used for Gaussianity tests is far less 
developed than its counterpart for power spectrum estimation.  
Even though Bayesianism has become the norm in current $C_\ell$
estimation, tests of Gaussianity have all, but one  \cite{mle}, been 
conducted in a frequentist vein. 
Most of these tests have revealed consistency with 
Gaussianity (e.g.\cite{kog96a}), but claims for non-Gaussianity 
detection in COBE-DMR 4-year maps have also 
been made\cite{fmg,pvl98,mag00}. Of these, one \cite{fmg}
has been found to be due to an experimental systematic \cite{band},
and another \cite{pvl98} to an error of method \cite{belen}.
The third claim \cite{mag00}, however, remains in place.  

In this paper we set up a proper Bayesian framework with
which to tackle the issue of non-Gaussianity. All Bayesian algorithms  
used so far have assumed that the signal, noise, and even 
galactic emissions are Gaussian random processes; and have targeted
the $C_\ell$ only. If the assumption  
of the signal's Gaussianity is to be dropped then one must propose an  
alternative form for the likelihood. Here the difficulties begin,
since a workable non-Gaussian likelihood cannot be easily defined.  
Non-Gaussianity spans an infinite dimensional space,
and truncations into a finite number of degrees of freedom are
usually inconsistent.

The only attempt made so far is published in \cite{mle}. In that work
the Edgeworth expansion \cite{kendall} was used to produce a Bayesian 
joint estimate 
of the $C_\ell$ and the skewness of the COBE-DMR 4-year maps. 
This approach is at best very approximate. Cumulants of distributions 
are subject to very complicated constraints \cite{kendall}, 
and setting them all 
to zero apart from one (in \cite{mle} the skewness)
is inconsistent with the basic properties of 
a distribution, such as positive definiteness or normalization. Furthermore 
the Edgeworth expansion is an asymptotic series, which is never
a distribution when truncated.

In this paper we bypass these difficulties by
deriving the general form of the likelihood from the  Hilbert space 
of a linear harmonic oscillator. We recall \cite{qm} that 
the ground state has a Gaussian wave-function, while all excited states 
have wave-functions which multiply a Gaussian by a Hermite  
polynomial. The space of all distributions can then be spanned by 
the amplitudes, $\alpha_n$, of the various energy eigenstates, 
with a general distribution 
taking the form of a Gaussian times the square of a (possibly 
finite) series  of Hermite polynomials.  

Such a generic distribution has a remarkable similarity with the
Edgeworth expansion, which takes the form of a Gaussian multiplying
an infinite series of Hermite polynomials with coefficients which
are themselves polynomials in the cumulants of the distribution. 
Closer comparison
of the two expressions reveals that indeed the 
amplitudes $\alpha_n$ can be written as series of cumulants\cite{joao}; 
these are the combinations of cumulants which can be varied 
independently. In particular these are the combinations which 
can be independently set to zero without mathematical inconsistency.
Furthermore, perturbatively (that is when the cumulants are ``small''
in a suitable sense), the amplitudes $\alpha_n$ are proportional to 
the $n^{th}$ order cumulant. In some sense the $\alpha_n$ generalize
cumulants to non-perturbative situations. 

We thus arrive at a well-defined mathematical framework for conducting
Bayesian tests of Gaussianity, which jointly produces $C_\ell$
estimates. Its interest is twofold. Firstly there is the obvious
interest in finding out whether the CMB fluctuations are Gaussian
or not. Secondly there is the issue of whether $C_\ell$ estimates
themselves may vary if non-Gaussian degrees of freedom are allowed
into the likelihood. In this paper we describe this formalism
(Section~\ref{meth}), and apply it to VSA simulations (Section~\ref{res}),
pending actual data. 

It is interesting to note that the $\alpha_n$ are more than 
a mathematical device, and have a physical interest of their
own within the framework of the standard inflationary
scenario\cite{guth81,linde,as,linde1}. Standard inflationary fluctuations
are Gaussian because the inflaton's fluctuations satisfy
harmonic oscillator equations, and are assumed to be in the ground state.
The latter is an {\it assumption} which  relies loosely on the 
boundary conditions imposed in quantum cosmology
\cite{hartle,jonathan,vilenkin}, and needs not be correct\cite{joao}.
A non-trivial wave-function for the inflaton's fluctuations
manifests itself in non-Gaussian density fluctuations, even 
if we do not depart from single-field, slow-roll, inflation. 
Quantifying their non-Gaussianity by means of the $\alpha_n$
offers a direct bridge to the wave-function of the inflaton's
fluctuations.

Hence, if we take it for granted that inflation is realized 
in its simplest form, and is triggered by quantum cosmology, 
then measuring the 
parameters $\alpha_n$ amounts to mapping the wave-function
of the Universe as it emerged out of a quantum epoch.

\section{Signal-to-noise eigenmodes}
\label{ston}

The non-Gaussian likelihood formalism we are about to present works most
simply when applied to a series of independent variables. We shall 
therefore combine our method with  the technique of signal-to-noise 
eigenmodes, which we start by reviewing. 

The signal-to-noise technique is a special case of the Karhunen-Loeve
method where the parameter dependence is linear (affine) 
\cite{bond94,bond98,tegmark96}. Let us consider a general 
set of random variables
\be\label{s-to-n1}
d_{i} = s_{i} + n_{i},
\ee
where $s_{i}$ is the signal component and $n_{i}$ is the noise 
counterpart. Let us also assume that the signal and noise contributions
are independent and each has zero mean.
The quantity $d_{i}$ could represent, for
example, the observed temperature fluctuation in the $i$th pixel of a
CMB map, or alternatively the real or imaginary part of the amplitude of the
$i$th coefficient in the Fourier expansion of the map. 
The covariance matrix $\langle d_i d_j \rangle$ is given simply by
$C_{ij} = \langle s_{i} s_{i'} \rangle + \langle n_{i} n_{i'} \rangle
\equiv S_{ij} + N_{ij}$, where $S$ and $N$ are the signal and noise
covariance matrices respectively.

In the standard likelihood approach one estimates the
parameters $a_1, a_2, \ldots, a_n$ of the probability distribution
from which the $s_i$ are drawn by calculating the likelihood 
${\cal L}({\mathbf d}| {\mathbf a})$ as a
function of these parameters. The parameters ${\mathbf a}$
usually enter the calculation through the signal covariance matrix 
$S({\mathbf a})$.
Let us suppose, in some part of the calculation, we are interested only
in the parameter $a_1$ (say), with the remaining parameters fixed at
particular values. In some (very common) cases, it proves possible to
write the signal covariance matrix in the special form
\be
S({\mathbf a}) = S_1(a_1) + S_2(a_2,\ldots,a_n),
\ee
where $S_1$ depends only on the parameter of interest and $S_2$
depends only on the (fixed) values of the other parameters and is thus
a fixed matrix. Furthermore, it is often also the case that $S_1$
is linear in $a_1$ and so can be written in the simple form 
$S_1(a_1) = a_1U$ where 
the fixed matrix $U = S_1(1)$. Thus the total covariance matrix $C$
can be written as
\be
C = S + N = a_1U + S_2 + N \equiv a_1U + V, \label{specialc}
\ee
where $U$ is the `unit signal' covariance matrix and $V$
is the `generalised noise' matrix. In particular, we note that if the
parameters ${\mathbf a}$ are the power spectrum coefficients $C_\ell$
(or averages of the $C_\ell$'s in given $\ell$-bins), it is always
possible to write the covariance matrix in the form (\ref{specialc}).

Since $U$ and $V$ are both real
symmetric matrices, they can be diagonalised simultaneously by a
single similarity transformation. This is most easily achieved by
solving the generalised eigenproblem $U{\mathbf x} 
= \lambda V{\mathbf x}$. Let us denote the
corresponding eigenvalues by $\lambda_i$ and eigenvectors by 
${\mathbf e}_i$,
which are normalised such that ${\mathbf e}_i^t V {\mathbf e}_i = 1$. 
If we now 
consider the new set of variables $\xi_i 
= {\mathbf e}_i \cdot {\mathbf d}$, then it is
straightforward to show that these new variables are uncorrelated 
for {\em any} value of the parameter $a_1$, with a 
covariance matrix given by
$\langle \xi_i \xi_j \rangle =
(1+a_1\lambda_i)\delta_{ij}$. The $\xi_i$  are the
so-called eigenmodes of signal-to noise ($S/N$); the
modes with large eigenvalue are expected to be well-measured 
while modes with small eigenvalues
are poorly-measured and do not contribute significantly to the
likelihood.

If the original data $d_i$ were distributed as a multivariate
Gaussian, the particular advantage of the $S/N$ eigenmode basis is
that the likelihood
function for the parameter $a_1$ (with $a_2,\ldots,a_n$ held fixed)
becomes a simple product of one-dimensional Gaussians, 
and can be written simply in terms of the eigenvalues
$\lambda_i$ as
\be
{\cal L}(\bxi|a_1)=
\prod_{i} \frac{1}{\sqrt{2\pi}\sqrt{1+a_1\lambda_i}} e^{-\frac{\xi_i^2}
{2(1+a_1\lambda_i)}}.  \label{gausslike}
\ee

It is clear that this procedure can be repeated for each of the other
parameters $a_2,\ldots,a_n$, provided in each case the covariance
matrix can be written in the form (\ref{specialc}). The
likelihood function can thus be evaluated very simply along the `coordinate
directions' in parameter space. Moreover, in the special case where the 
parameters $a_1,a_2,\ldots,a_n$ are
mutually independent (or quasi-independent to a good approximation),
the likelihood function factorises as
\be
{\cal L}({\mathbf d}|{\mathbf a})
= {\cal L}({\mathbf d}|a_1){\cal L}({\mathbf d}|a_2)\cdots 
{\cal L}({\mathbf d}|a_n). \label{lfactors}
\ee
Thus, in this case, the above signal-to-noise eigenmode procedure can
be repeated for each parameter in turn to evaluate the full likelihood
function.

\section{An exact, non-perturbative, non-Gaussian likelihood}
\label{meth}
Let $x$ represent a general random variable, within a set of variables
which are assumed to be independent. Let us build its
distribution from the space of wave-functions which are 
energy eigenmodes of a simple harmonic oscillator. The following
results may be found in any quantum mechanics book (e.g.\cite{qm})
adopting the Schrodinger (rather than the Heisenberg) picture.
We have that the general wave-function for $x$ is given by:
\be\label{psiexp} 
\psi(x)=\sum_n \alpha_n\psi_n (x)
\ee 
where $n$ labels the energy spectrum $E=\hbar \omega (n+1/2)$.
The basis functions $\psi_n$ take the form 
\be 
\psi_n(x)=C_nH_n{\left(x\over 
{\sqrt 2} \sigma_0\right)}e^{-{x^2\over 4 \sigma_0^2}} 
\ee 
with normalisation fixing $C_n$ as, 
\be 
C_n={1\over (2^n n!{\sqrt{2\pi}}\sigma_0)^{1/2} }\, . 
\ee 
The only constraint upon the amplitudes $\alpha_n$ is:
\be\label{const}     
\sum_n |\alpha_n|^2 =1
\ee
This is 
a simple algebraic expression which can be eliminated explicitly
by writing:
\be\label{elim}
\alpha_0= \sqrt{1- \sum_1^\infty |\alpha_n|^2 }
\ee
The quantity
$\sigma_0^2$ is the variance  associated with the (Gaussian) probability 
distribution for the ground state $|\psi_0|^2$.  We shall work with Hermite 
polynomials $H_n(x)$ defined as  
\be\label{herm} 
	H_n(x) = (-1)^n e^{x^2} \frac{d^x}{d x^n}e^{- x^2} 
\ee 
and normalised as 
\be\label{ortho} 
\int^\infty_{-\infty} e^{-x^2}H_n(x)H_m(x)dx=2^n \pi^{1/2} n!\,\,\delta_{nm}. 
\ee 
The most general probability density for the fluctuations  
in $x$ is thus:
\be 
P=|\psi|^2={e^{-{x^2\over 2 \sigma_0^2}}}
\left|\sum_n \alpha_n C_n H_n{\left(x\over {\sqrt 2}\sigma_0\right)}\right|^2
\ee

The ground state (or ``zero-point'') fluctuations are Gaussian, 
but any admixture with other states will be reflected 
in a non-Gaussian distribution function. Accordingly we may use
the amplitudes of these admixtures, $\alpha_n$, as non-Gaussianity
indicators. Their obvious advantage is the rather trivial constraint
(\ref{const}), which can be ignored using (\ref{elim}). It permits
concentrating on a finite set of non-Gaussian degrees of freedom,
without mathematical inconsistency. 

However there is another reason why the $\alpha_n$ are of mathematical
interest: they reduce to cumulants $\kappa_n$ under certain assumptions.
If we  assume mild non-Gaussianity (which we define 
through the condition  $|\alpha_0|^2\gg |\alpha_i|^2$, for $i\ge1$)
then to first order in $\alpha_i$, 
\be\label{expan}
P(x)=|\psi|^2={e^{-{x^2\over 2 \sigma_0^2}}\over {\sqrt{2\pi}} 
\sigma_0}{\left[1+\sum_{n\ge 1}{2\Re (\alpha_n)\over (2^n n!)^{1/2}} 
H_n{\left(x\over {\sqrt 2}\sigma_0\right)}\right]} 
\ee 
where we have taken $\alpha_0$ to have zero phase (so that to 
first order $\alpha_0=1$). Comparing (\ref{expan})
with the Edgeworth expansion \cite{kendall}
we find a one-to-one correspondence between the amplitudes 
of the various energy eigenstates, and the combinations of cumulants 
appearing as coefficients in the Edgeworth expansion. The latter simplify 
enormously if we only keep first order terms, that is if we assume
that quadratic and higher order terms in the cumulants are negligible.
Then we find that 
\be
\kappa_n\propto \Re(\alpha_n)
\ee
with a rather complicated proportionality constant (which is easy
to work out case by case). Hence, 
to first order, the coherent contamination of the ground state by  
the $n$\th energy eigenstate is signalled by a non-vanishing cumulant 
$\kappa_n$. For instance 
the presence of the third energy eigenstate results in  
$\kappa_3\propto \Re (\alpha_3)\ne 0$, 
and, to first order, zero higher order cumulants.  
 
The advantage of using the $\alpha_n$ is that they still work
(i.e. they still lead to proper distributions) when the distribution
is highly non-Gaussian. Any maximum likelihood estimate will necessarily
have to probe regions of $\alpha_i$ beyond the perturbative regime,
even for Gaussian realizations. In these regions  setting all
but a finite number of  $\kappa_n$ to zero is inconsistent; but
not for a finite number of $\alpha_n$. In the non-perturbative regime
the $\alpha_n$ become rather complicated series of $\kappa_n$.
However these series of cumulants may be varied, or set to zero,
independently,
and still lead to a distribution. Hence we should regard the $\alpha_n$
as non-perturbative generalizations of cumulants. 

More concretely, in the non-perturbative regime, we have 
\be  
P(x)=|\psi|^2={e^{-{x^2\over 2 \sigma_0^2}}\over {\sqrt{2\pi}} 
\sigma_0}\sum_{i,j}{\alpha_i^*\alpha_j\over (2^{i+j} i!j!)^{1/2}} 
H_i{\left(x\over {\sqrt 2}\sigma_0\right)} 
H_j{\left(x\over {\sqrt 2}\sigma_0\right)} 
\ee 
We may recover the Edgeworth expansion by noting that,  
\be\label{expand} 
    e^{-x^2} H_i(x)H_j(x)  =  
e^{-x^2} \left[ \sum_n b^{ij}_n H_n(x)\right] 
\ee 
with, 
\be\label{bij} 
	b^{ij}_n =  
\frac{2^{s-n}i!j!}{(s-n)!(s-i)!(s-j)!} 
\ee 
with $2s=n+i+j$. One may derive (\ref{bij}) 
using (\ref{ortho}) and the standard result for the 
integral over a product of three Hermite polynomials (formula 7.375.2  
of \cite{Grad}). Thus we obtain the more complicated expression 
\be  
P(x)=|\psi|^2={e^{-{x^2\over 2 \sigma_0^2}}\over {\sqrt{2\pi}} 
\sigma_0}\sum_n{\left(\sum_{i,j}{b^{ij}_n 
\alpha_i^*\alpha_j\over (2^{i+j} i!j!)^{1/2}}\right)} 
H_n{\left(x\over {\sqrt 2}\sigma_0\right)} 
\ee 
Comparing with the Edgeworth expansion leads to the rather complicated
non-perturbative expression 
relating the amplitudes $\alpha_i$ and series of cumulants.

\section{Application to VSA simulations}
\label{res}

We have applied our method to simulated observations by the Very Small Array
(VSA) interferometer. The VSA has been built by Cambridge and Jodrell
Bank in the UK, and is sited at the Teide Observatory in Tenerife. It has
just become operational. The VSA is expected to give detailed maps of
the CMB anisotropy with a sensitivity $ \sim 5 \mu K$ and covering a
range of angular scales from $10'$ to $2^{\circ}$ for a frequency
range of 28 and 38 GHz. It will have 14 antennas and a 2-GHz bandwidth
analogue correlator and uses the same technology as
Cosmic Anisotropy Telescope (CAT). The VSA is able to observe in
`compact' and `extended' modes, which are sensitive to different
$\ell$-ranges. In the compact mode, it is expected that
VSA will recover the angular power spectrum in 10
spectral bins where each bin
is centred respectively at $\ell \approx 114$, 211, 308, 404, 501, 598, 695,
792, 889, 986. The width of each bin $\Delta\ell \approx 97$
corresponds to the $1/e$ diameter of the aperture function of the 
interferometer and represents the length scale on which the underlying
Fourier modes of the sky are correlated by the instrument. Thus, the
spectral bins have been chosen so that the power spectrum 
estimates in each bin are quasi-independent to a good approximation.

An account of a maximum-likelihood method for analysing interferometer
observations of the CMB anisotropies is given in \cite{hobson95}. In
the standard likelihood analysis the parameters of interest are 
$a_k = \langle \ell^2 C_\ell/(2\pi) \rangle_{\mbox{{\small $k$th bin}}}$. Since
these parameters are quasi-independent,
the full likelihood function factorises as in (\ref{lfactors}) to a
good approximation. The
signal-to-noise eigenmode procedure outlined in section \ref{ston} is
then applied to each factor, which, in
the standard approach, reduces to the simple product of
one-dimensional Gaussians given in (\ref{gausslike}). 

In our new approach, however, we subject the standard 
Gaussian likelihood algorithm to
the following modifications.  Instead of assuming the simple Gaussian
form for the probability distribution of each S/N eigenmode $\xi_i$, 
we instead consider the more general situation in which all
$\alpha_n$ are set to zero, except for the real part of
$\alpha_3$. The reason for this is that such a quantity reduces to the
skewness in the perturbative regime. The imaginary part of $\alpha_3$
is only meaningful in the non-perturbative regime (and can be set to
zero independently without inconsistency). Hence we are considering a
likelihood of the form:
\be\label{like}
P(x)= \frac{e^{\frac{-x^{2}}{2 \sigma_{0}^{2}}}}{\sqrt{2 \pi}
\sigma_{0}} 
\left[ \alpha_{0} + \frac{\alpha_{3}}{\sqrt{48}} 
H_{3} \left(\frac{x}{\sqrt{2} \sigma_{0}}\right) \right]^{2} 
\ee
with 
\be
\alpha_0= \sqrt{1- \alpha_3^{2}}
\ee
explicitly replaced in (\ref{like}).

The generalization of this distribution to the multidimensional
case is trivial in the signal-to-noise eigenmode basis, since we can
simply take the product of the individual one-dimensional distributions.
Thus, when considering the power spectrum in the $k$th spectral bin,
we adopt the likelihood function
\be
{\cal L}(\bxi|a_k,\alpha_3)
=\prod_{i} \frac{e^{-\frac{\xi_{i}^{2}}{2(1+a_k\lambda_i)}}}   
{\sqrt{2\pi}\sqrt{1+a_k\lambda_i}} 
\left[ \alpha_{0} + \frac{\alpha_{3}}{\sqrt{48}} 
H_{3}\left(\frac{\xi_{i}}{\sqrt{2 (1+a_k\lambda_i)}}\right)\right]^{2}, 
\ee
where $a_k$ is the average of value $\ell^2 C_\ell/(2\pi)$ in the
$k$th spectral bin. The $\alpha_3$ could in principle depend
on $\ell$, but for simplicity we have dropped this dependence.
Notice that because the noise is Gaussian, and because of the 
principle of superposition in quantum mechanics, the $\alpha_n$
appearing in this formula are the ones pertaining to the signal
alone.

We have applied this method to a $30\times 12$ hour simulated VSA observation
of a Gaussian CMB realisation drawn from a standard inflationary model 
with $\Omega_{\rm cdm}=0.95$, $\Omega_{\rm b}=0.05$, 
$\Omega_{\Lambda}=0$, $h=0.5$, $n_s=1$ and no tensor
contribution. In Fig.~1 we show the contour plots of the likelihood functions for
the amplitude of the power spectrum $\ell^{2} C_\ell/(2\pi)$ and 
$\alpha_3$ for each of the 10 spectrals bins observed by VSA. 
The alignment of the contour axes with the
coordinate axes implies the reassuring result that there seems to be
little correlation in each bin between the power spectrum estimate 
and the estimate of the non-Gaussian parameter.

In Fig.~2 we plot the likelihood function for $\alpha_3$ in each spectral bin
after marginalization over the amplitude of the angular power
spectrum.
The results obtained indicate that the value of $\alpha_3$ scatter around 
$\alpha_3=0$, within the range implied by the width of the likelihood.
The percentage of the population inside the contour
intersecting the origin represents the confidence level for
rejecting Gaussianity. All of these are within 1-2 sigma,
indicating that this method is not biased. Notice that for those
bins in which we failed to obtain a CMB detection (see Fig.3)
there seems to be a bias towards a peak at $\alpha_3=0$, without
the scatter expected from the width of the likelihood.

In Fig.~3 we plot the likelihood functions for the amplitude
of the power spectrum in each bin after marginalization over the 
parameter $\alpha_3$ (solid line).
Superimposed are the corresponding conditional distributions for $\alpha_{3}=0$ (dashed line). 
For this Gaussian CMB realisation, 
the distributions obtained are not significantly 
affected by the inclusion of the extra parameter $\alpha_3$.
The most noticeable effect is a slight variation of the position of the peak (particularly for bins 1 and 2) which is in agreement with Fig~1.
Since in each bin the estimate of the power spectrum and the
non-Gaussian parameter are weakly correlated, 
we see that the widths of the likelihood functions for the 
angular power spectrum are not significantly increased by including 
the $\alpha_{3}$ parameter.

In Fig.~4 we plot the joint likelihood for $\alpha_3$ obtained by multiplying the individual likelihoods for the 10 spectral bins, thus obtaining the overall estimate and a better constraint on $\alpha_{3}$.  

Finally, in Fig~5 we plot the distribution of the peak of the likelihood for $\alpha_{3}$ obtained from Monte Carlo simulations. In each VSA simulation the CMB is a realization drawn from an inflationary model. The CMB fluctuations are thus Gaussian.
The distribution peaks around a value of $\alpha_{3}=0$ confirming that our algorithm is indeed not biased.

\section{Conclusions}
In this paper we laid down the foundations for a rigorous
Bayesian framework for testing non-Gaussianity, and jointly
estimating the power spectrum (Section~\ref{meth}).
Our main achievement was to convert testing
Gaussianity into a problem of Bayesian estimation. We defined
a series of parameters $\alpha_n$, to be added to the power spectrum,
such that the origin of the new space  represents Gaussianity. 
These parameters are generalizations of cumulants. If all cumulants are very
small, in a suitable sense, each of the new parameters is approximately 
proportional to a cumulant. If not, then the new parameters become 
series of powers of cumulants. They are desirable non-perturbative
generalizations of cumulants because they are independent, ie:
subject to essentially no constraints, unlike standard cumulants. 

With any dataset, one must then determine the contour of the
likelihood intersecting the origin of the $\alpha_n$ space,
after marginalization over the power spectrum. 
The percentage of the population inside this contour is the
confidence level for rejecting Gaussianity. We found that 
for simulated VSA observations of a Gaussian CMB realisation this
confidence level is always within 1-2 sigma. 
To assess if our algorith is unbiased one must produce simulated VSA observations of several Gaussian CMB realizations. We found that the distribution of the peak of the likelihood in $\alpha_{3}$ for a number of these realizations peaks around a value of $\alpha_{3}=0$ showing that our method is indeed not biased.
(Section~\ref{res}).

The method we have proposed is completely general, and may be
applied to any type of experiment, interferometric or single-dish. 
In particular its application to COBE-DMR maps, closely mimicking 
the steps of \cite{cont}, is straightforward. The only issue which 
may complicate the method is galactic foreground removal. In some
experiments foregrounds away from the galactic plane may be ignored,
by suitably choosing the frequency channels. In some cases,
contaminations may be removed by subtracting off the correlated
component, making use of templates \cite{kog96b}. In these
cases there is no extra complication to our method. 

However in some cases \cite{hobson95}
foreground subtraction is part of the maximum
likelihood algorithm leading to CMB $C_\ell$ estimates. 
In some of these cases it is assumed that Galactic foregrounds
form a Gaussian random field. With our method we may now
allow for non-Gaussian degrees of freedom to be applied to
these emissions. Hence we should be able to improve significantly
on these methods of foreground removal, as well as exploring
signal non-Gaussianity. The detailed implementation of this
project, as well as its application to VSA data, is 
the subject of a future publication. 

The formalism we have developed is also of assistance
for generating realizations belonging to the most general
ensemble parameterized by the $\alpha_n$. In work in preparation
we show how this can be done, and how the maximum likelihood
method proposed in this paper may then differentiate between
distinct distributions on the basis of single realizations. 

\newpage

\begin{figure}
\begin{center}
\resizebox{7in}{!}{\includegraphics{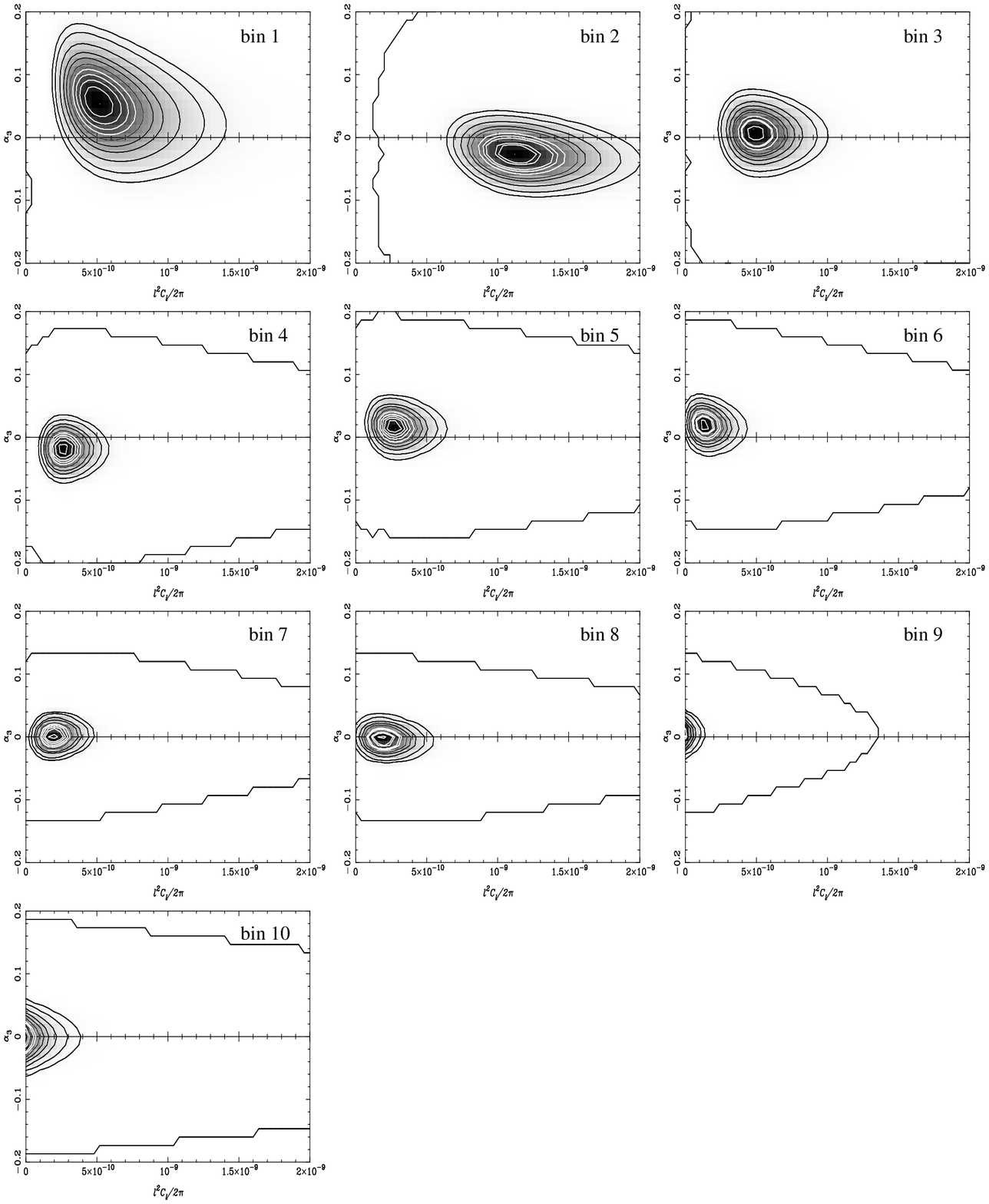}}
\caption{Confidence contours for a simulated VSA observation of a
Gaussian CMB realisation drawn from a standard inflationary model.
In each spectral bin the fluctuations are parametrized by the average 
value of $l^{2}C_{l}/(2\pi)$ and the generalised cumulant $\alpha_{3}$. 
Contours are at 10,20,30,...90 and
95 per cent confidence levels.
\label{fig:1}}
\end{center}
\end{figure}

\newpage

\begin{figure}
\begin{center}
\resizebox{6in}{!}{\includegraphics{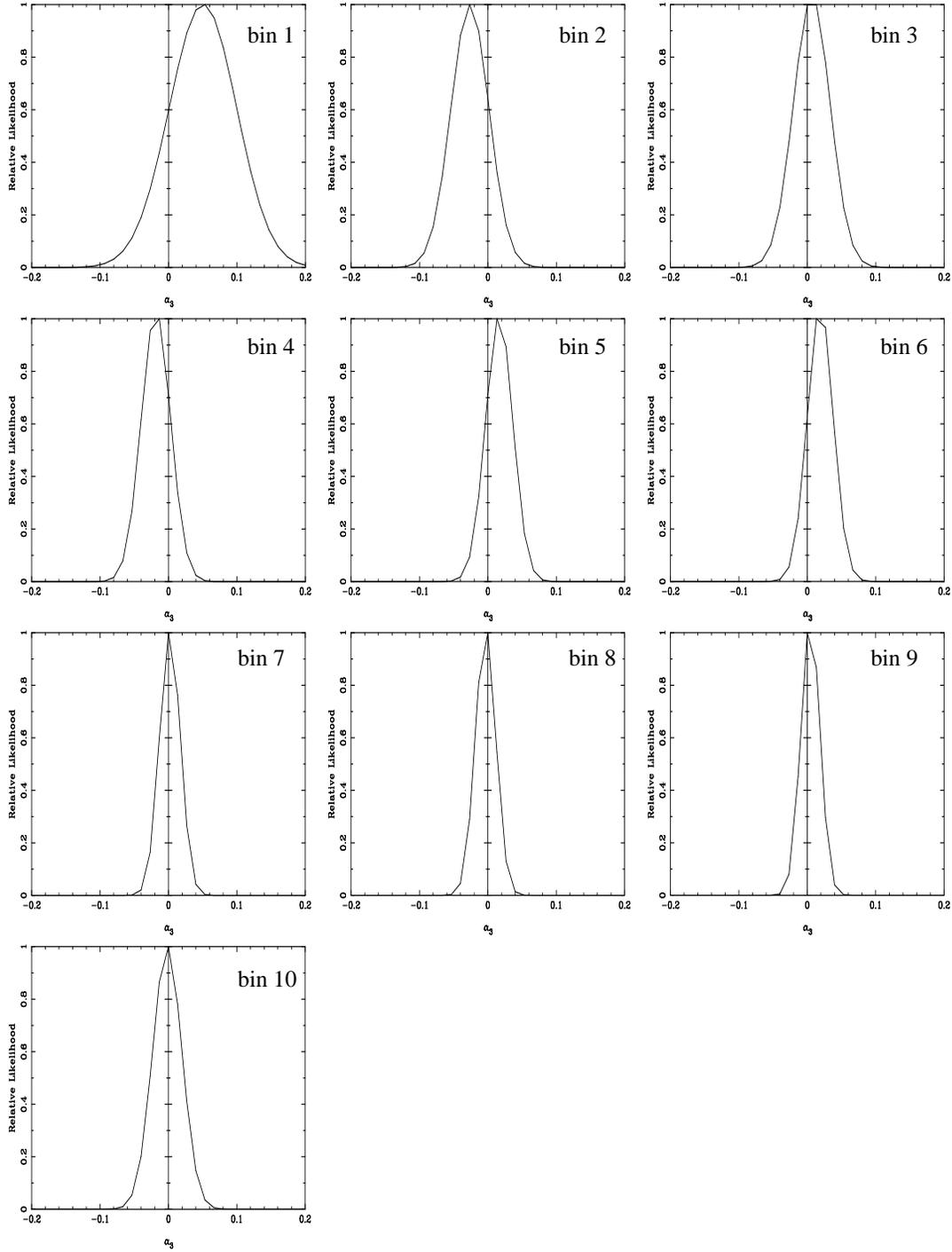}}
\caption{Marginalised likelihood functions for $\alpha_{3}$ in 
each spectral bin obtained from the simulated VSA observation.
\label{fig:2}}
\end{center}
\end{figure}

\newpage

\begin{figure}
\begin{center}
\hspace{0.4in}
\vbox{%
\hbox{%
\resizebox{2in}{2in}{\includegraphics{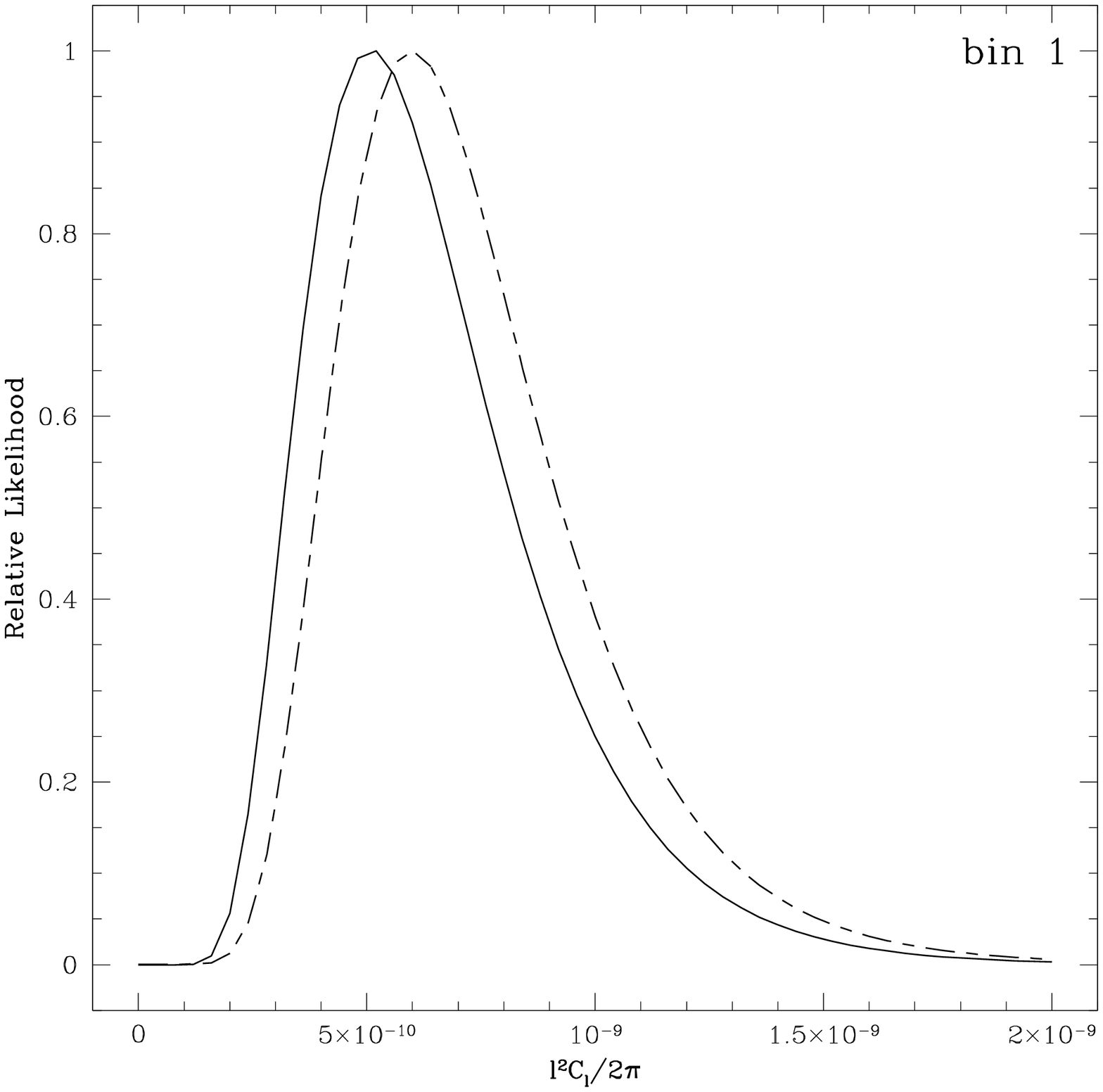}}
\resizebox{2in}{2in}{\includegraphics{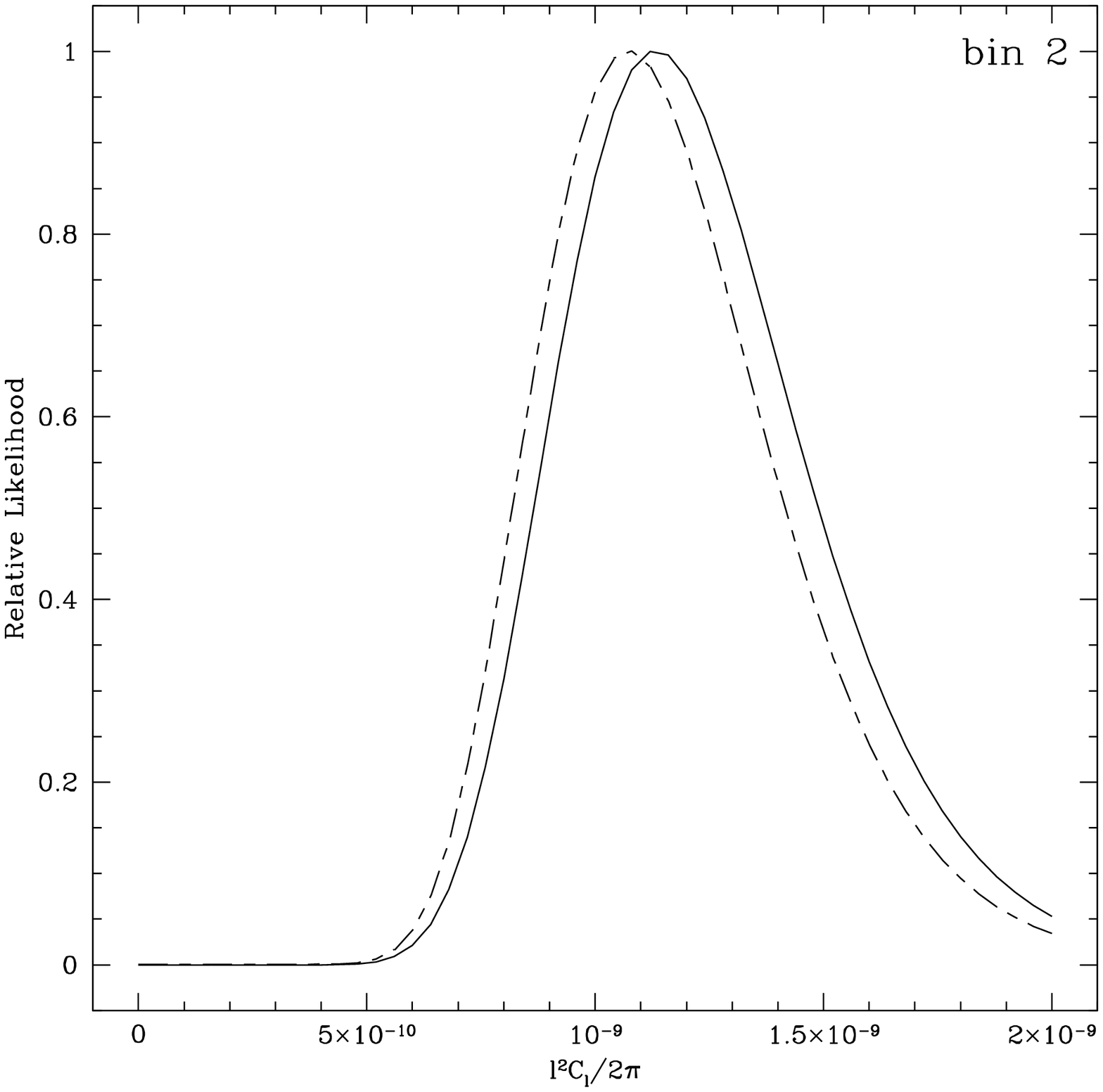}}
\resizebox{2in}{2in}{\includegraphics{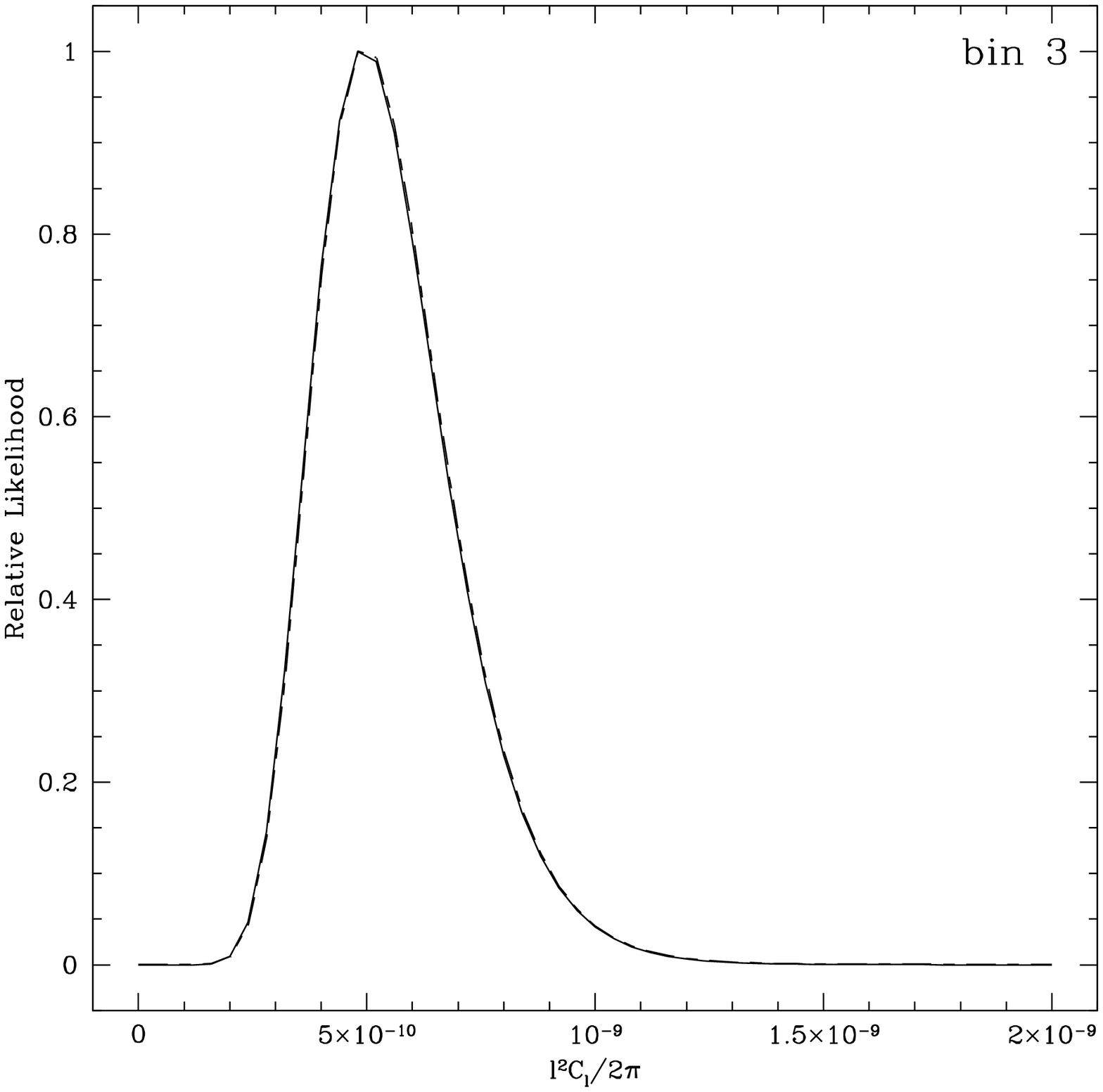}}
}
\hbox{%
\resizebox{2in}{2in}{\includegraphics{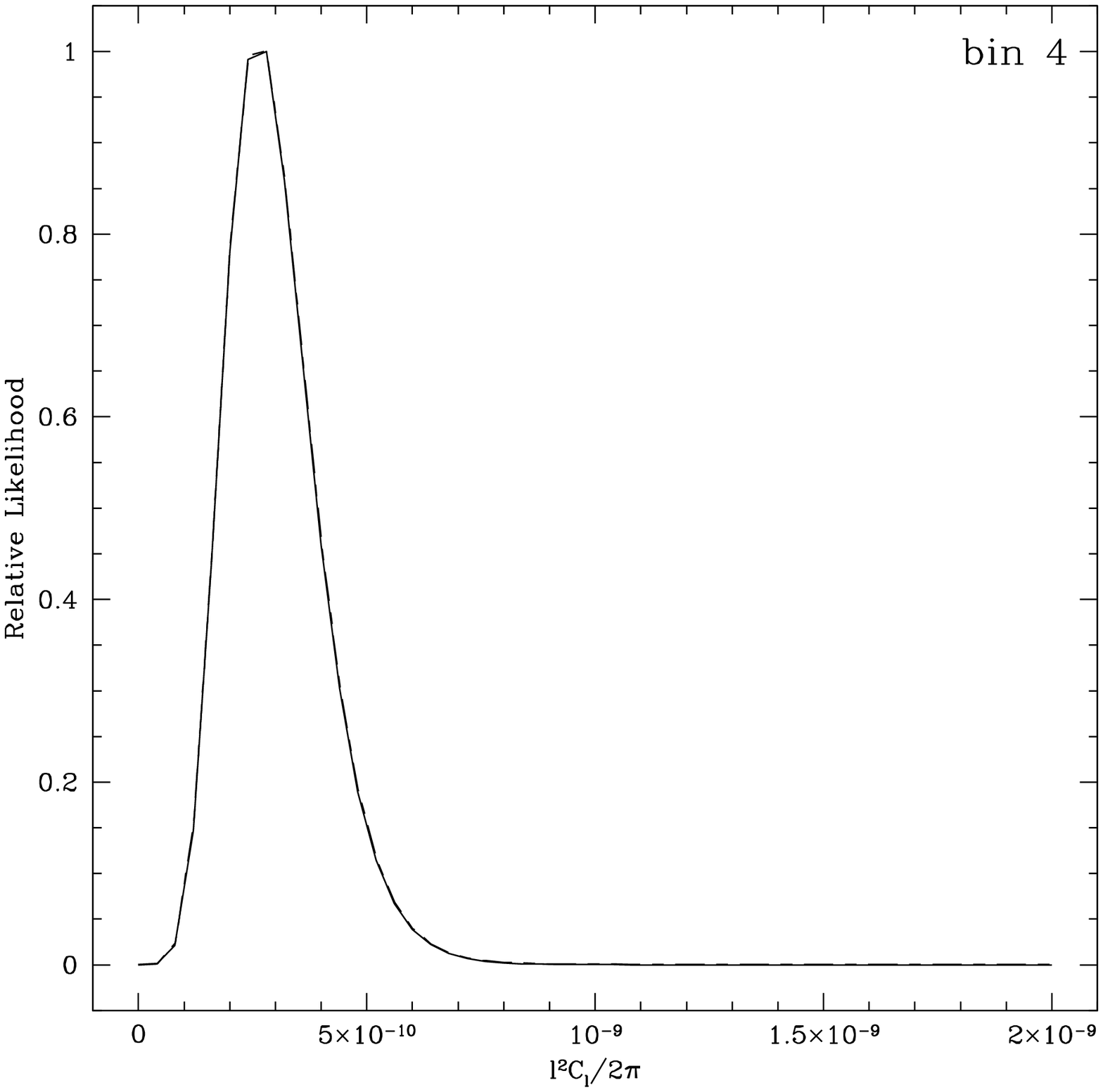}}
\resizebox{2in}{2in}{\includegraphics{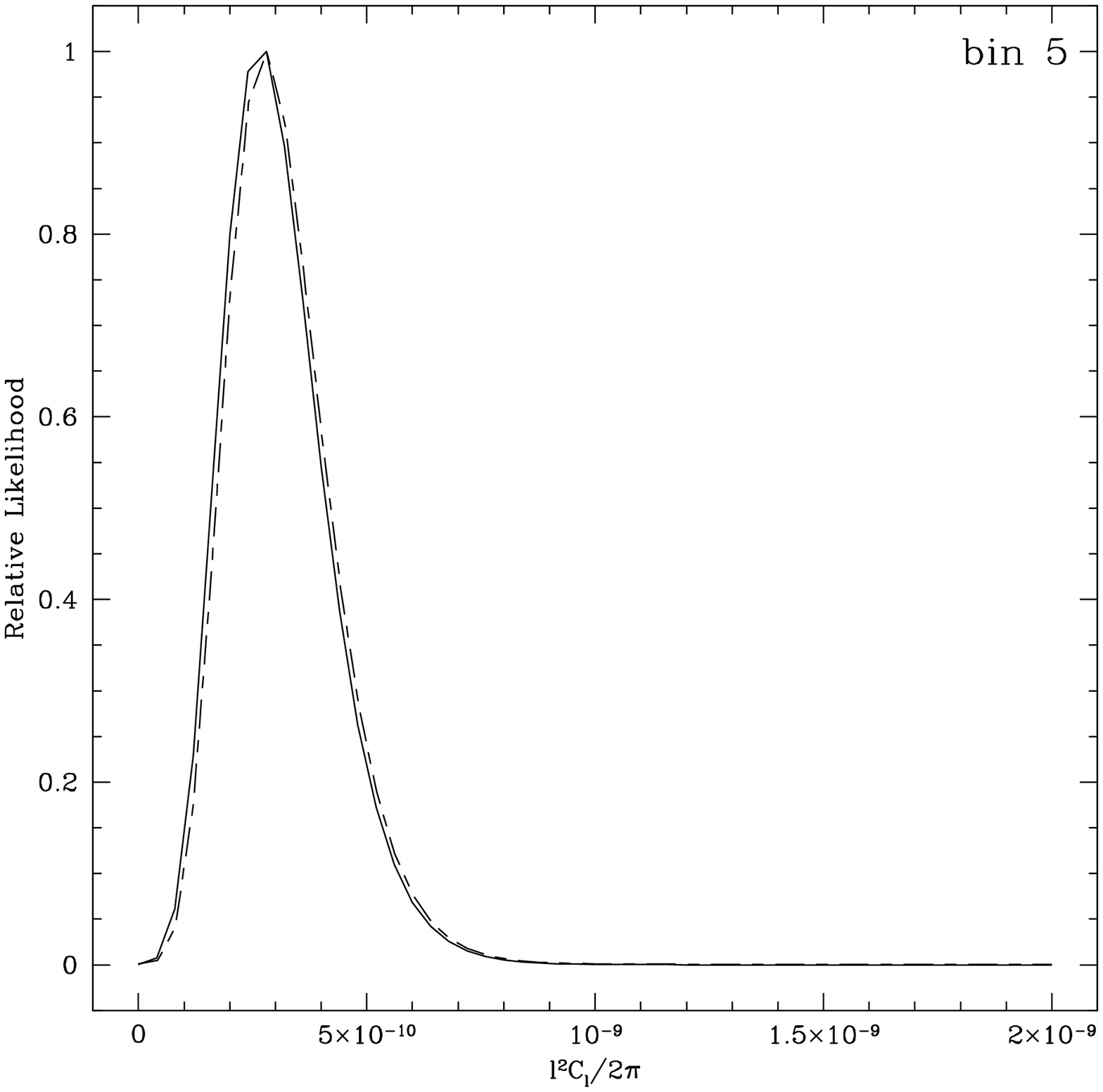}}
\resizebox{2in}{2in}{\includegraphics{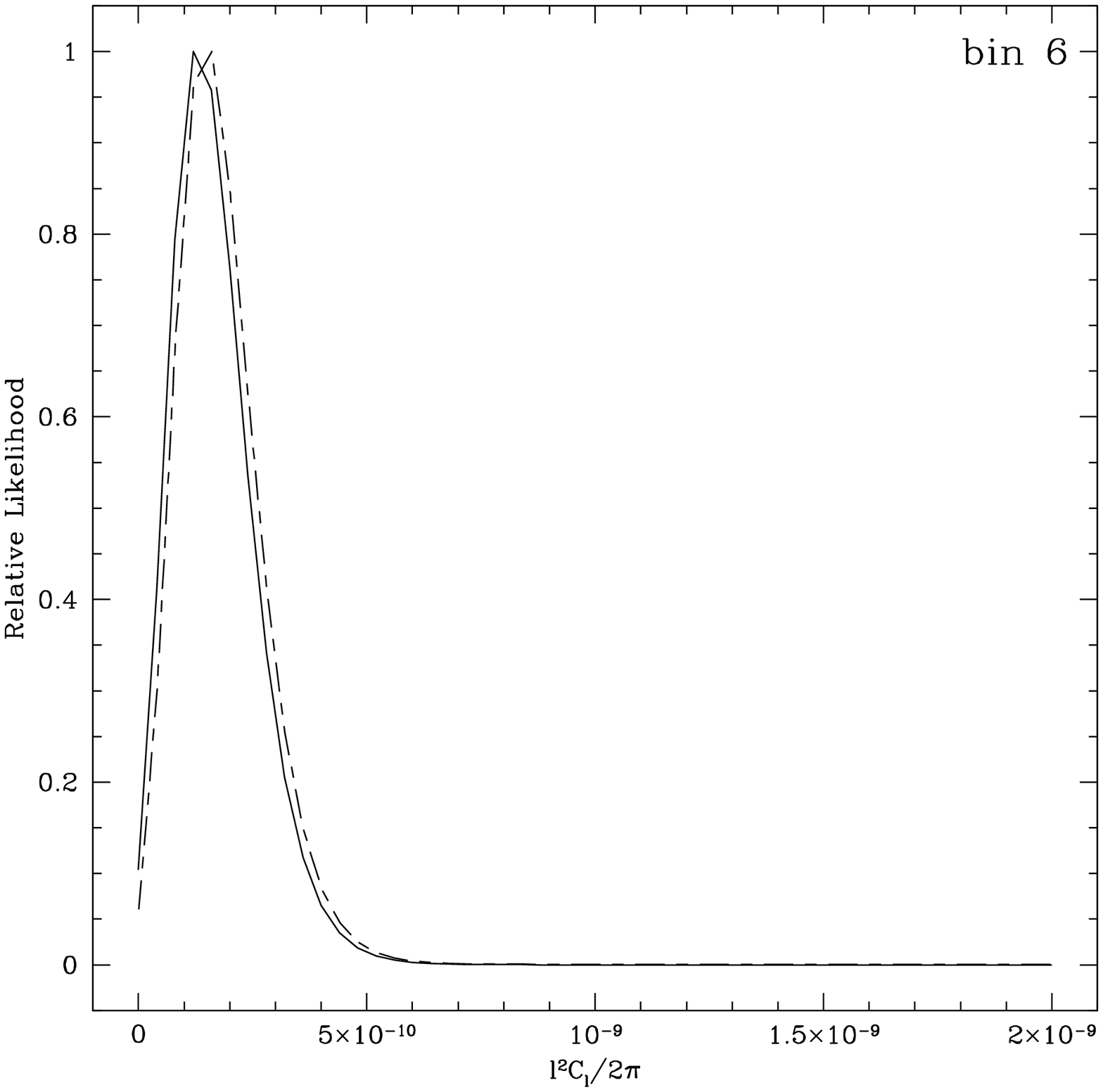}}
}
\hbox{%
\resizebox{2in}{2in}{\includegraphics{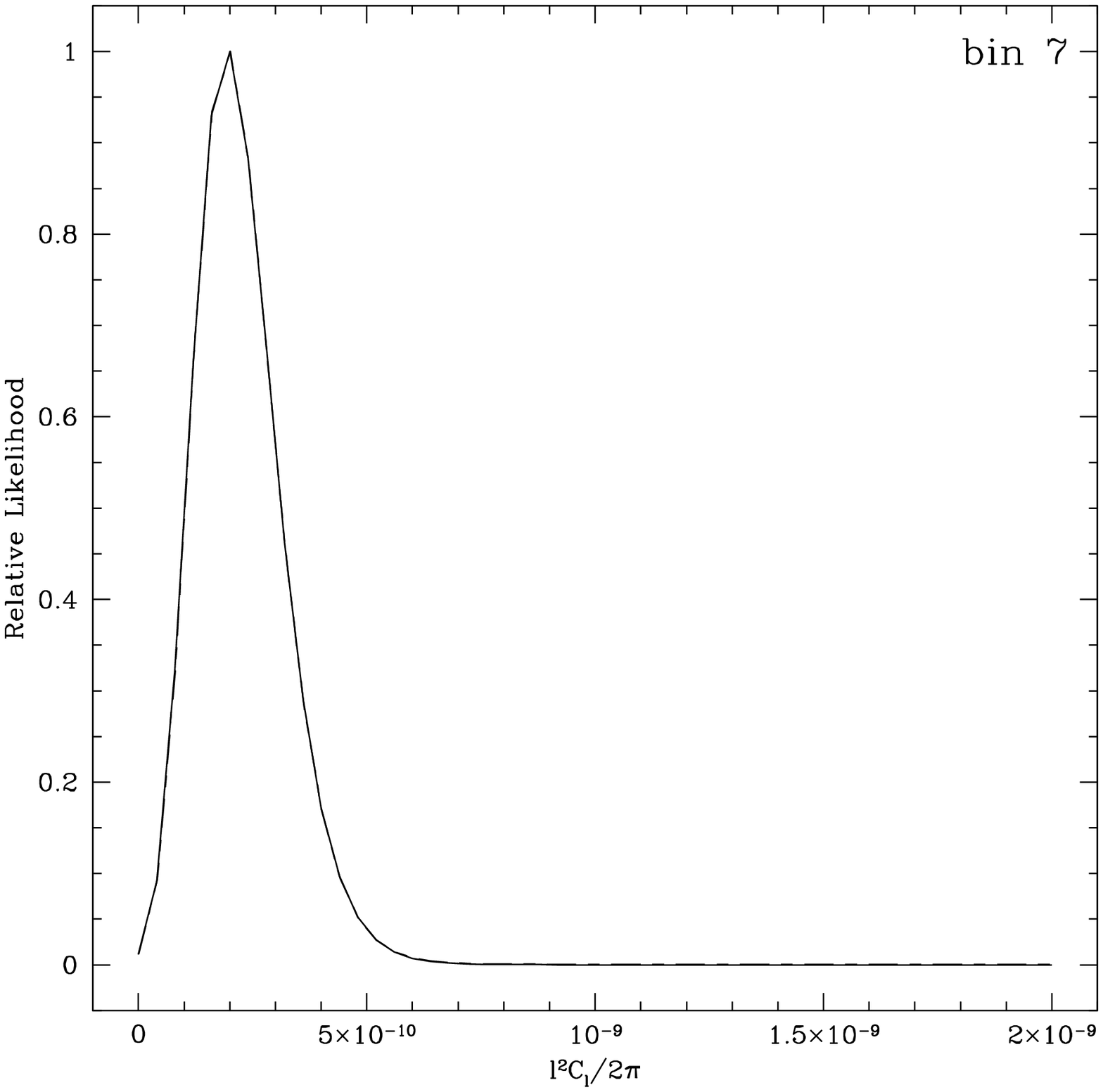}}
\resizebox{2in}{2in}{\includegraphics{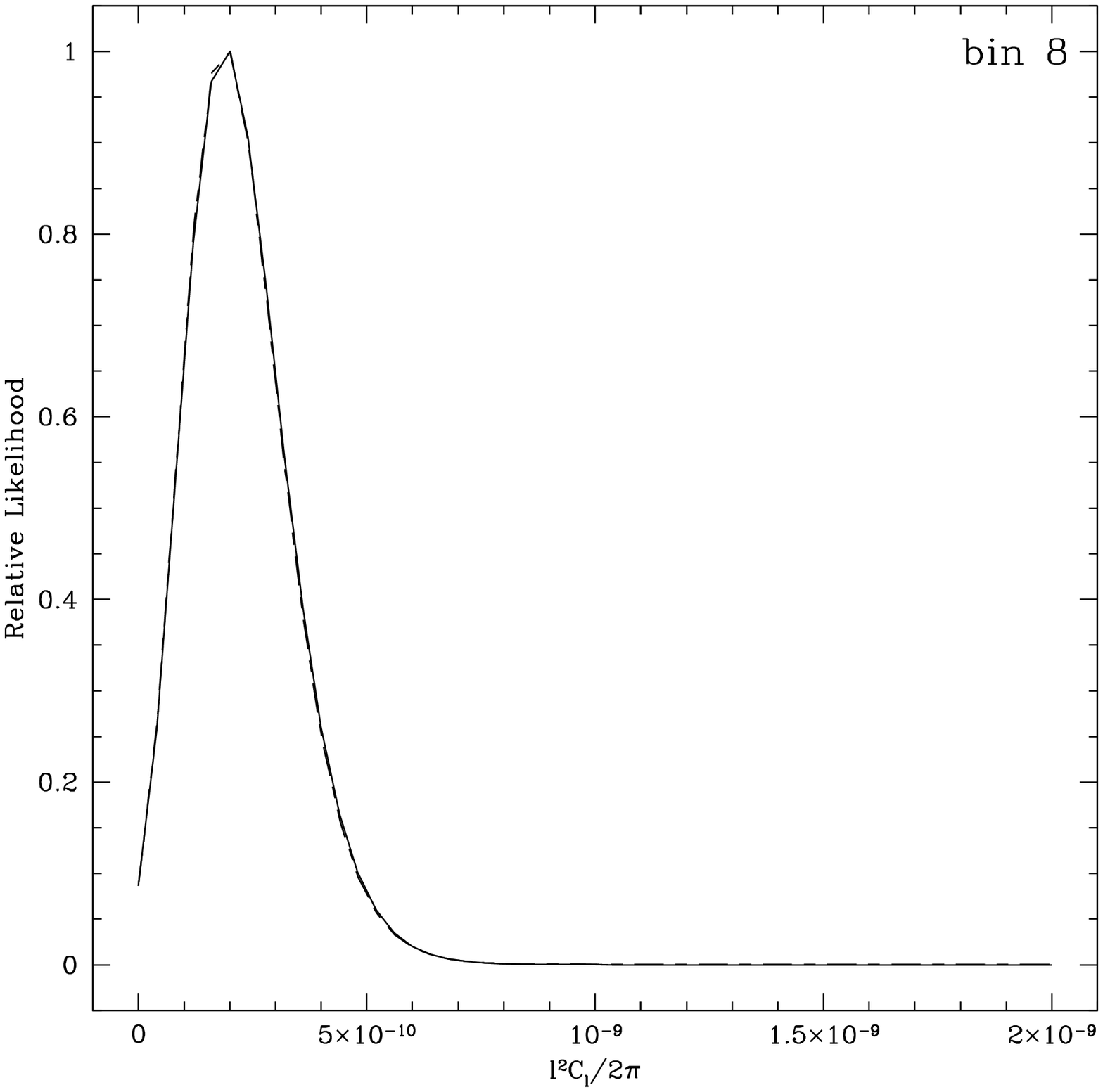}}
\resizebox{2in}{2in}{\includegraphics{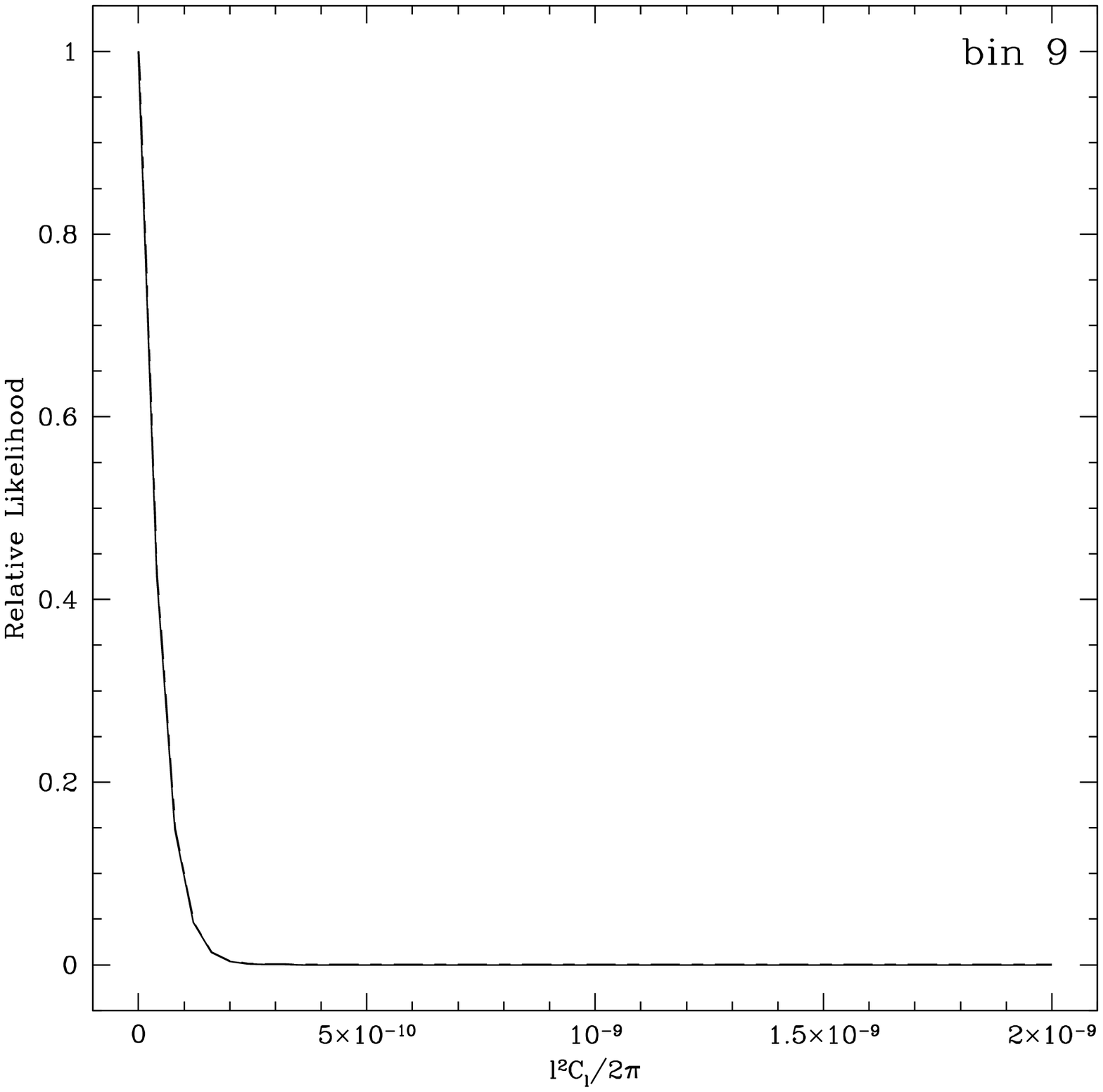}}
}
\hbox{%
\resizebox{2in}{2in}{\includegraphics{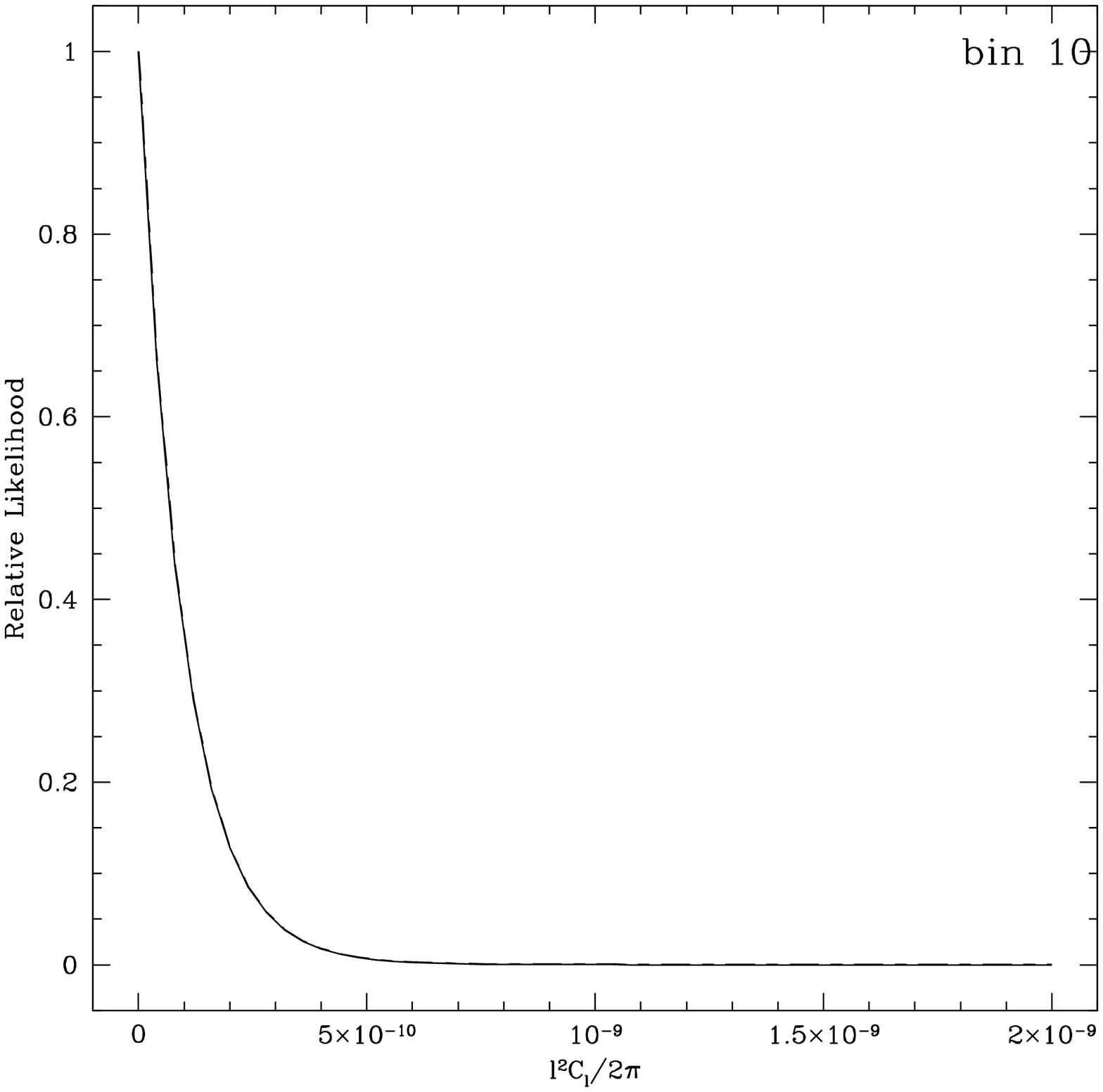}}
}
}
\caption{Solid line: marginalized distribution for CMB fluctuations parameterized by $l^{2}C_{l}/(2\pi)$, for a simulated VSA observation; dashed line: conditional distribution for $\alpha_{3}=0$.
\label{fig:3}}
\end{center}
\end{figure}

\newpage

\begin{figure}
\centering
\resizebox{2.5in}{2.5in}{\includegraphics{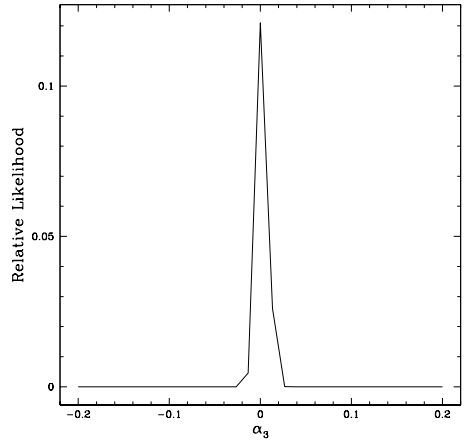}}
\caption{The joint likelihood for $\alpha_3$ obtained by multiplying the individual likelihoods in $\alpha_{3}$ for the 10 spectral bins.}
\end{figure}
\begin{figure}
\centering
\resizebox{2.5in}{2.5in}{\includegraphics{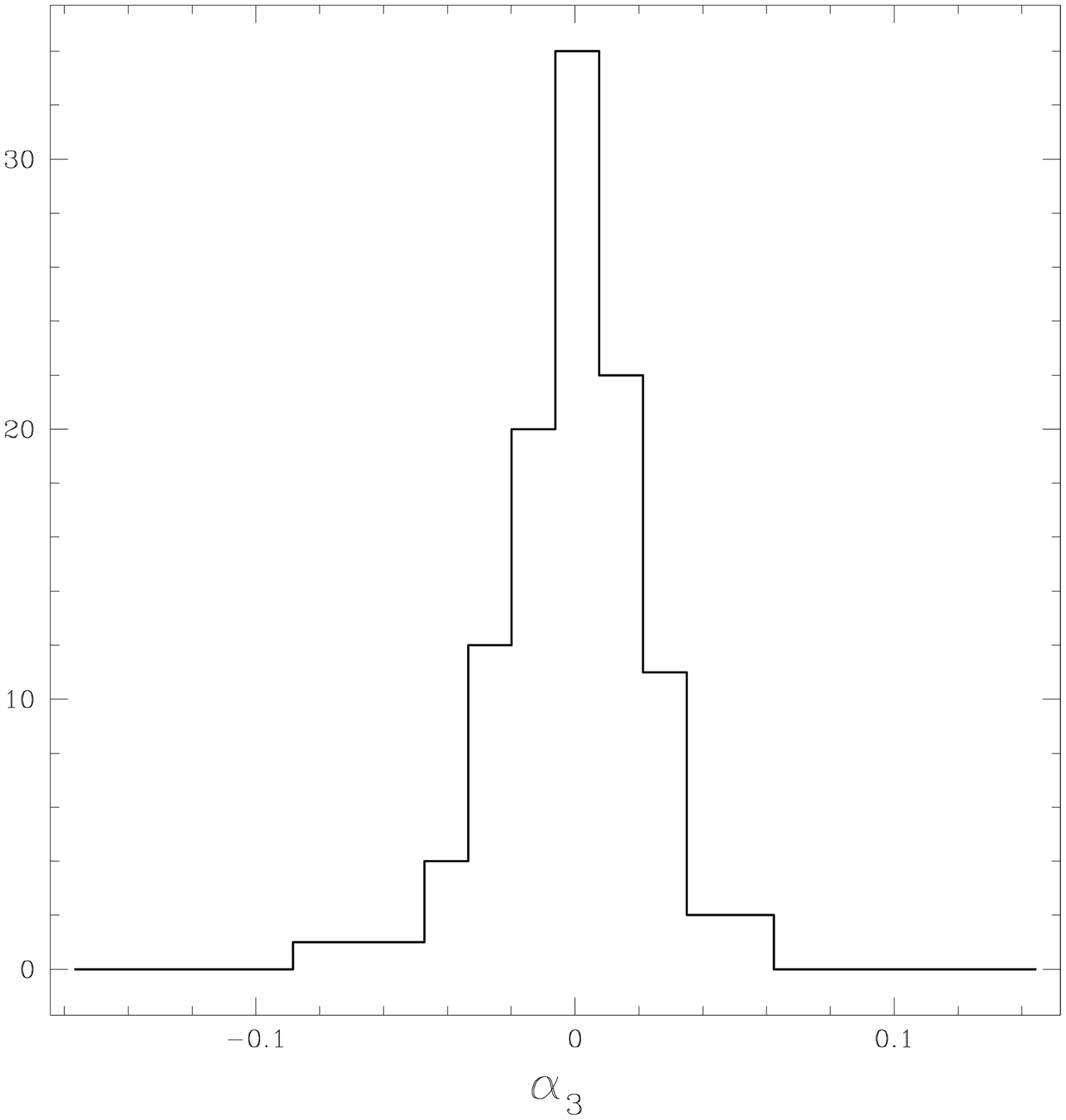}}
\caption{The distribution of the peak of the likelihood in $\alpha_{3}$ for simulated VSA observations of several Gaussian CMB realizations.}
\end{figure}

\section*{Acknowledgments}
JM would like to thank Rachel Bean and Carlo Contaldi for 
inspiring this paper during the preparation of \cite{joao}.
JM thanks the Isaac Newton Institute for support and hospitality 
at the initial stages of this project.
GR would like to thank Pedro Ferreira for enlightening discussions. 
GR also 
thanks the Dept. of Physics of the University of Oxford for support 
and hospitality during the progression of this work. 
GR is funded by FCT (Portugal) under `Programa PRAXIS XXI', 
grant no. PRAXIS XXI/BPD/9990/96. 
MH acknowledges funding from PPARC in form of an Advanced Fellowship.

\end{document}